\documentclass[onecolumn,amsmath,amssymb]{revtex4}
\usepackage{graphicx}
\begin{document}

\title
      [Chameleon Dark Energy]
      {Chameleon Dark Energy}

\author{Ph. Brax}
\affiliation{Service de Physique Th\'eorique, CEA-Saclay, Gif/Yvette
  cedex, France F-91191}

\author{C. van de Bruck}
\affiliation{Astro-Particle Theory and Cosmology Group, Department of
  Applied Mathematics, University of Sheffield, Sheffield S3 7RH, UK}

\author{A.C. Davis}
\affiliation{DAMTP, Centre for Mathematical Sciences, Cambridge
  University, Wilbeforce Road, Cambridge CB3 0WA, UK}

\author{J. Khoury}
\affiliation{Center for Theoretical Physics, Massachusetts Institute of Technology,
 Cambridge, MA, 02139, USA}

\author{A. Weltman}{
\affiliation{Institute for Strings, Cosmology and Astroparticle Physics,
  Columbia University, New York, NY, 10027, USA}


\begin{abstract}
Chameleons are scalar fields whose mass depends on the environment, specifically on the ambient matter density.
While nearly massless in the cosmos, where the matter density is tiny,
their mass is of order of an inverse millimeter on Earth, where the density is high.
In this note, we review how chameleons can satisfy
current experimental constraints on deviations from General Relativity (GR).
Moreover, we study the cosmological evolution with a chameleon field
and show the existence of an attractor solution, akin to the tracker solution
in quintessence models. We discuss how chameleons can naturally drive the observed acceleration
of the universe.
\end{abstract}

\maketitle


Dark energy is one of the most puzzling aspect of our observed
Universe.
The lack of a theoretical explanation for an almost vanishing
cosmological constant has motivated a host of phenomenological studies where
the dynamics of the vacuum energy are modeled with a scalar field,
usually dubbed quintessence. Most of these models have a long time attractor solution, thereby eradicating the problem of sensitivity to initial conditions for a rolling scalar field.
The existence of an attractor solution requires the scalar potential to be of the runaway form, as in the inverse power-law potential of the Ratra-Peebles model.
In order for the equation of state of the dark energy to be observationally distinguishable from that of a cosmological constant, the field must be rolling today, which in turns requires its mass to be of order
of the present Hubble rate.
Thus the quintessence field acts as a massless field whose
interactions with ordinary matter are therefore tightly constrained
by current bounds on violations of the Equivalence Principle (EP).
This conflicts with general expectations from string theory, where scalar fields typically arise as the shape or size of extra dimensions and therefore couple to matter with gravitational strength.

Chameleon scalar fields~\cite{cham,cham2,cham3,cham4} provide an alternative mechanism for circumventing the constraints from local tests of gravity. In these models, the scalar field acquires a mass which depends on the ambient matter density. In the cosmos, where the density is minuscule, its mass can be on the order of the Hubble constant, allowing the field to be rolling on cosmological time scales. On Earth, however, where the density is many orders of magnitude higher, the chameleon acquires a mass that is sufficiently large to satisfy all current experimental bounds on deviations from GR.

To see how this works, consider the following general scalar--tensor theory:
\begin{equation}
S =\int d^4 x\sqrt{-g} \left\{\frac{M_{Pl}^2R}{2}- \frac{(\partial \phi)^2}{2} -V(\phi)
+ {\cal L}_m(\psi_m,A^2(\phi)g_{\mu\nu})\right\}
\end{equation}
where $\phi$ is the chameleon scalar field with scalar potential $V(\phi)$, assumed to be of the runaway form as in general quintessence models. See Fig.~\ref{fig1}. Fermion (matter) fields, denoted by $\psi_m$, couple conformally to the chameleon through the $A^2(\phi)$ dependence of the matter lagrangian ${\cal L}_m$.

This conformal coupling leads to an extra term in the Klein-Gordon equation for the chameleon, as usual proportional to the trace of the matter stress-tensor:
\begin{equation}
\nabla^2 \phi = V_{,\phi} -  \alpha_\phi T^\mu_\mu\,,
\label{KG1}
\end{equation}
where $\alpha_\phi\equiv \frac{\partial \ln A}{\partial \phi}$.
With the approximation that the matter is well described by a pressureless (non-relativistic) perfect fluid with density $\rho_m$, this reduces to
\begin{equation}
\nabla^2 \phi = V_{,\phi} + \alpha_\phi\rho_m A(\phi)\,,
\label{KG2}
\end{equation}
where $\rho_m$ is conserved with respect to the Einstein frame metric $g_{\mu\nu}$.

An immediate realization is that the dynamics of $\phi$ are governed by an effective potential, $V_{{\rm eff}}$, which depends explicitly on $\rho_m$:
\begin{equation}
V_{\rm eff}(\phi)=V(\phi) +\rho_m A(\phi)\,.
\end{equation}
While the ``bare'' potential $V(\phi)$ is of the runaway form, the effective potential
will have a minimum if $A(\phi)$ increases with $\phi$. This is shown in Fig.~\ref{fig1}.
Moreover, the location of this minimum and the mass of small fluctuations, $m^2=V_{,\phi\phi}^{{\rm eff}}$, both depend on $\rho_m$.
In other words, the physical properties of this field vary with the environment, thus the name chameleon.

\begin{figure}
\includegraphics[width=5 in] {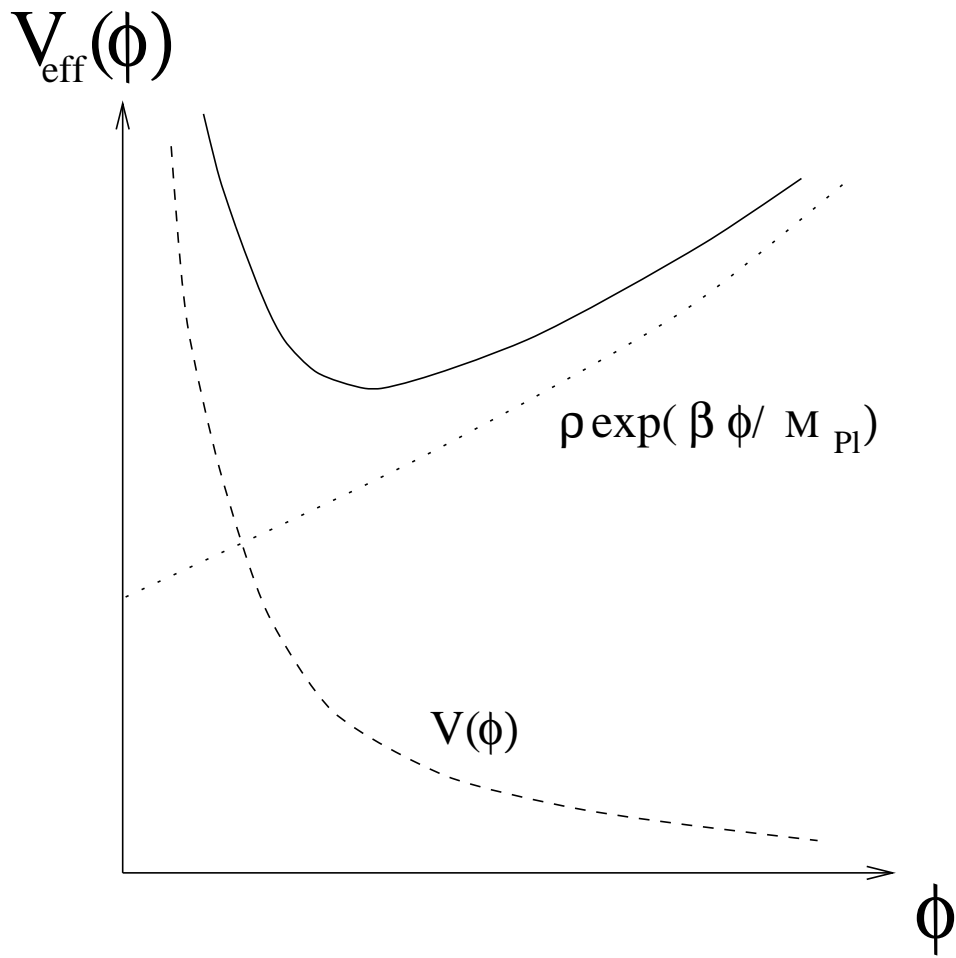}
\caption{The effective potential for the chameleon (solid line) is the sum of the ``bare'' potential, $V(\phi)$, which is of the runaway form (dashed line), and a density-dependent term (dotted line). Here we choose the exponential coupling: $A(\phi) = e^{\beta\phi/M_{Pl}}$}
\label{fig1}
\end{figure}

Due to its direct coupling to matter, the chameleon mediates a ``fifth force'' which is therefore subject to a host of experimental constraints.
Two important effects suppress the magnitude of this force.
Firstly, the mass of the chameleon is an increasing function of the density. Thus, in regions of high density, such as on Earth, the range of the chameleon can be less than 1~mm, which is sufficiently short to satisfy current bounds from fifth force searches in the laboratory.
Nevertheless, its range can in principle be large in the cosmos, where the density is many orders of magnitude smaller, thereby allowing the chameleon to be a rolling quintessence field driving the current phase of cosmic acceleration.
Hence, the large hierarchy between the local and cosmic densities translates into a large hierarchy in the corresponding chameleon ranges.

The second effect is the so-called ``thin-shell'' mechanism. For sufficiently large objects, the $\phi$-force on a test particle is almost entirely due to a thin shell of matter just below the surface of the object, while the matter in the core of the object contributes negligibly. In other words, only a small fraction of the total mass of the object affects the motion of a test particle outside. This breakdown of the superposition principle is a direct consequence of the non-linearity of Eq.~(\ref{KG2}).

To illustrate this, we derive an approximate solution for the chameleon for a spherically-symmetric object of radius $R$ and homogeneous density $\rho$. For concreteness, we focus on the inverse power-law potential, $V(\phi) = M^{4+n}/\phi^n$, where $M$ has units of mass, and an exponential coupling, $A(\phi)= e^{\beta\phi}$, of gravitational strength, $\beta=O(1)$. The relevant boundary conditions for this problem are that the solution be non-singular at the origin and that $\phi$ tends to its ambient value, $\phi_0$, far from the object.

For sufficiently large objects, one finds that, within the object, the field assumes a value $\phi_c$ which minimizes the effective potential: $V_{,\phi}(\phi_c) + \beta\rho_c e^{\beta\phi_c/M_{Pl}}/M_{Pl} = 0$. This holds true everywhere inside the object except within a thin shell of thickness $\Delta R$ below the surface where the
field grows. Outside the object, the profile for
$\phi$ is essentially that of a massive scalar, $\phi \sim
\exp(-m_0r)/r$, where $m_0$ is the mass of the chameleon in the ambient medium.

The thickness of the shell is related to $\phi_0$, $\phi_c$,
and the Newtonian potential of the object, $\Phi_N=M/8\pi M_{Pl}^2R$, by
\begin{equation}
\frac{\Delta R}{R} \approx \frac{\phi_\infty-\phi_c}{6\beta
M_{Pl}\Phi_N}\,.
\label{DR}
\end{equation}
The exterior solution can then be written explicitly as~\cite{cham}
\begin{equation}
\phi(r)\approx -\left(\frac{\beta}{4\pi M_{Pl}}\right)\left(\frac{3\Delta
R}{R}\right)\frac{M e^{-m_0 (r-R)}}{r} + \phi_0\,.
\label{thinsoln}
\end{equation}

Evidently, this derivation only makes sense if the shell is thin: $\Delta R/R \ll 1$.
Keeping everything else fixed, we see from Eq.~(\ref{DR}) that this is the case for objects with sufficiently large $\Phi_N$. Then Eq.~(\ref{thinsoln}) says that the correction to Newton's law at short distances is given by $F= (1+\theta) F_N$, where
\begin{equation}
\theta = 2\beta^2 \left(\frac{3\Delta R}{R}\right) \,,
\end{equation}
which is small. Hence a thin shell guarantees a small deviation from Newton's law.

Fifth force experiments are usually performed inside a vacuum chamber where the density is negligibly small. Inside a cavity of radius $R_{cav}$, explicit calculations~\cite{cham} show that the chameleon assumes a nearly constant value $\phi_0$, with $\phi_0$ satisfying $m_0 R_{cav} \sim {\cal O}(1)$. That is, the interaction range of the chameleon-mediated force inside the cavity is of order of the size of the cavity.

Thus the two test masses used to measure $G$ must have a thin shell, for otherwise the correction to Newton's constant $G_N$ from the chameleon-mediated force will be of order unity. Since $G_N$ is known to an accuracy of $10^{-3}$, they must satisfy
\begin{equation}
\frac{\Delta R}{R} = \frac{\phi_0-\phi_c}{6\beta
M_{Pl}\Phi_N} \;\sim\; 10^{-3}\,.
\end{equation}
where we assume that $\beta=O(1)$. 
For typical test masses and cavity of characteristic radii of $\sim 1$~cm and $\sim 1$~m, respectively, this gives $\phi_0 \sim 10^{-28}\;M_{Pl}$. For the inverse power-law potential, $V(\phi) = M^{4+n}/\phi^n$, with $n\sim {\cal O}(1)$, this translates into a constraint on $M$:
\begin{equation}
M\sim 10^{-3}\  \hbox{eV}\,.
\label{Mcond}
\end{equation}
The fact that the upper bound coincides with the mass scale of the vacuum energy today is remarkable and unexpected since it was derived purely from considerations of local tests of gravity, independently of any cosmological input. We shall exploit this coincidence below to explain the observed cosmic acceleration. We note in passing that Eq.~(\ref{Mcond}) implies a chameleon range of $ 0.1 $~mm in the atmosphere, thereby ensuring that violations of the EP are exponentially small locally.

In space, however, the range is much longer, of order $10-1000$~AU. Although this is long range, orbits of planets and satellites are nevertheless unaffected since these large bodies all have a thin shell~\cite{cham}. On the other hand, two test masses aboard a satellite will in general not have a thin shell. This is because the ambient value of the chameleon is much larger in space than on Earth, making the thin shell condition much harder to satisfy in orbit. See Eq.~(\ref{DR}). It follows that the chameleon predicts a correction to Newton's law between these two test masses given by
\begin{equation}
\theta=2\beta^2\,,
\end{equation}
which is of order unity! Similarly, violations of the EP in space can be much larger than current bounds from the laboratory. These striking predictions will be tested by forthcoming satellite experiments, such as SEE, $\mu$SCOPE, GG and STEP.

Equation~(\ref{Mcond}) says that $M=10^{-3}$~eV is allowed by local tests of gravity. We exploit this coincidence to construct chameleon models leading to an accelerating universe today~\cite{cham2}. Consider a family of potentials involving a single mass scale $M$:
\begin{equation}
V=M^4 f(\phi/M)\,,
\end{equation}
where $f$ leads to ordinary quintessence with a long time tracking
solution. A typical example is $f(x)=e^{1/x^n}$.
Since $\phi \gg M$ ($x\gg 1$) both locally and on cosmological scales today, 
one has $V\approx M^4 + M^{4+n}/\phi^n$ in this case. Thus, cosmologically, this mimics a cosmological constant of the right magnitude to cause the universe to accelerate today. For tests of gravity, however, the constant piece is negligible, and the potential reduces to the Ratra-Peebles model considered earlier.

In a cosmological context, Eq.~(\ref{KG1}) reduces to
\begin{equation}
\ddot{\phi} + 3H\dot{\phi} = -V_{,\phi} - \frac{\beta}{M_{Pl}}\rho_m e^{\beta\phi/M_{Pl}}\,,
\end{equation}
where $\rho_m$ is the energy density in non-relativistic (dust) component. Note that we have neglected the contribution to $T^\mu_\mu$ from relativistic degrees of freedom, an assumption we will soon revisit.

As the universe expands, the matter density redshifts as usual as $a^{-3}$, implying that $V_{eff}$ evolves in time. To be more precise, the field value at the minimum, $\phi_{min}$, increases with time, as seen from Fig.~\ref{fig1}. However, it can be shown that the mass of small fluctuations about the minimum satisfies $m \gg H$. That is, the characteristic response time of the chameleon, $m^{-1}$, is much shorter than the time over which the potential evolves, $H^{-1}$, and the evolution is therefore adiabatic. In other words, if the chameleon starts at the minimum, it then stays at the minimum as the latter evolves in time. Being at the minimum is therefore a dynamical attractor.

More general initial conditions fall into {\it undershoot} and {\it overshoot} solutions, analogous to their counterparts in quintessence models. Undershoot solutions correspond to the chameleon starting from a value $\phi_i\gg \phi_{min}$. In this regime, the bare potential can be neglected, and one has
\begin{equation}
\ddot{\phi} + 3H\dot{\phi} \approx \frac{\beta}{M_{Pl}}T^\mu_\mu\,,
\end{equation}
where we have reintroduced $T^\mu_\mu$. As mentioned earlier, it is common to approximate $T^\mu_\mu\approx 0$ during the radiation-dominated (RD) era since a relativistic fluid has zero trace. If this were the case, the field would remain frozen at $\phi_i$ due to Hubble damping and would not reach the attractor early enough. Fortunately, however, in a realistic context the trace is not always negligible during the RD era. Indeed, as the universe expands and cools, different particle species successively become non-relativistic. Whenever this happens, $T^\mu_\mu$ becomes non-zero for about one e-fold of expansion and thus drives the chameleon towards the minimum. In other words, the chameleon is subject to a series of kicks, each bringing it closer to $\phi_{min}$. The cumulative effect of these kicks is a total field displacement of~\cite{cham2}
\begin{equation}
|\Delta\phi|\approx \beta M_{Pl}\,.
\label{Dphi}
\end{equation}
This is illustrated in Fig.~\ref{fig2}.
As we will see below, the chameleon must be at the minimum by the onset of big bang nucleosynthesis, otherwise it will result in an unnacceptably large time variation in masses and couplings. 
Therefore, from Eq.~(\ref{Dphi}), this is satisfied as long as $\phi_i \sim M_{Pl}$ for $\beta\sim {\cal O}(1)$.
Once the chameleon reaches the vicinity of the minimum, it oscillates and quickly settles down to the minimum.

\begin{figure}
\includegraphics[width=5 in] {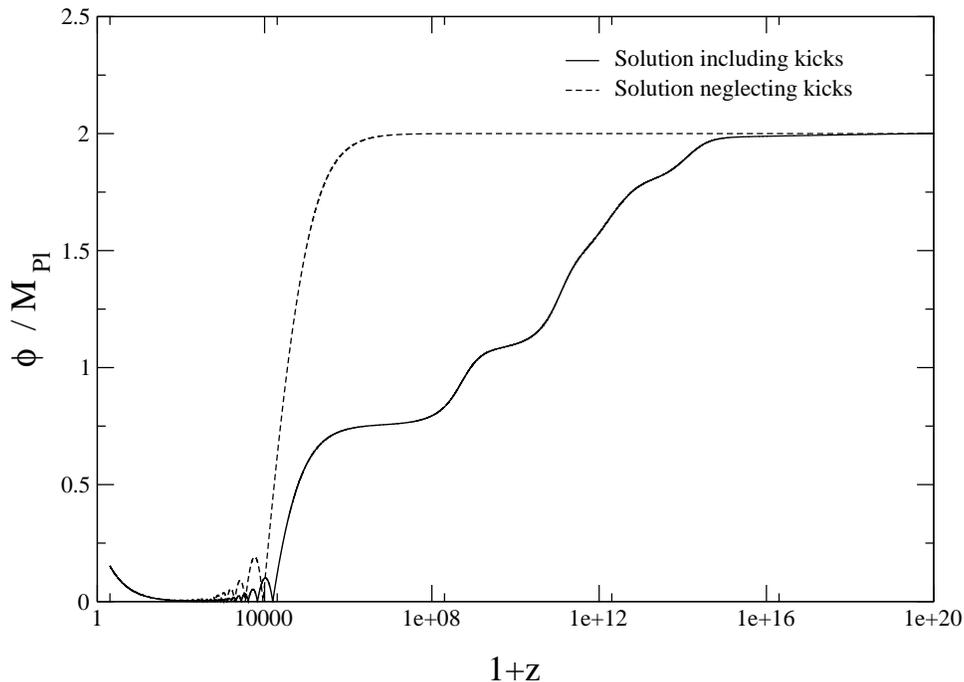}
\caption{The effect of kicks on the chameleon. Ignoring the kicks (dashed curve),
 the chameleon remains frozen at its initial value during the RD era due to Hubble damping. 
Including the kicks (solid curve), however, results in a total displacement of order $M_{Pl}$.}
\label{fig2}
\end{figure}

The overshoot solution corresponds to $\phi_i\ll\phi_{min}$. In this case, the field starts high up on the potential and quickly becomes kinetic-dominated. It overshoots the minimum and eventually comes to a halt due to Hubble damping at some value $\phi_{stop}\gg \phi_{min}$. It is easy to show that the latter is given by $\phi_{stop}= \phi_i + \sqrt{6\Omega_\phi^{(i)}}M_{Pl}$, where $\Omega^{(i)}_\phi$ is the initial chameleon fractional energy density. Afterwards, the solution is the same as in the undershoot case. In particular, the constraint from BBN requires $\phi_{stop}  \sim M_{Pl}$, and thus
\begin{equation}
\Omega_\phi^{(i)} \sim 1/6\,,
\label{Om}
\end{equation}
which, in particular, is consistent with equipartition at reheating.

We now explain how Eq.~(\ref{Om}) follows from BBN constraints.
Due to its coupling to matter, a variations of $\Delta\phi$ in the chameleon changes particle masses by
\begin{equation}
\left \vert \frac{\Delta m}{m}\right \vert = \beta \frac{\vert \Delta \phi\vert}{M_{Pl}}\,.
\end{equation}
The measured abundance of light elements constrains $|\Delta m|/m$ to be less than about 10\% or so from the time of BBN until today. In other words, BBN requires $|\phi_{BBN} - \phi^{(0)}|\sim 0.1\;M_{Pl}$ for $\beta\sim {\cal O}(1)$, where $\phi_{BBN}$ and $\phi^{(0)}$ denote the field values at BBN and today, respectively. Fortunately, at the minimum, the chameleon satisfies $\phi_{min}\ll M_{Pl}$ for all relevant times. Thus this bound is easily satisfied if the chameleon is at the minimum by the onset of BBN. However, if the field is far from the minimum at BBN, then the electron kick, which becomes important during BBN, will displace it by about $0.3\;M_{Pl}$, and therefore violate this bound. Therefore, it is crucial that the field be settled at the minimum by the onset of BBN. As mentioned above, this will be the case if Eq.~(\ref{Om}) is satisfied.


In conclusion, we have studied both the gravitational consequences and
the cosmology of chameleon models. Despite being strongly coupled to
matter, chameleon fields can act as a dark energy component accounting for the observed acceleration of the universe,
while satisfying all current experimental constraints on deviations from GR.



\end{document}